\documentclass[twocolumn]{aastex631}

\def\lapprox{$_<\atop{^\sim}$}

\usepackage{CJK}
\usepackage{bm} 
\usepackage{amsmath}
\usepackage{multirow}
\let\tablenum\relax
\let\tablewidth\relax
\usepackage{siunitx}
\usepackage{tabularray}
\usepackage{xcolor}

\newcommand{\Secref}[1]{\hyperref[#1]{Section~\ref*{#1}}}
\newcommand{\Appref}[1]{\hyperref[#1]{Appendix~\ref*{#1}}}

\shorttitle{Search for exo-satellites around GQ Lup B}
\shortauthors{Horstman et al.}

\graphicspath{{./}{figures/}}

\begin{document}
\begin{CJK*}{UTF8}{gbsn}

\title{RV measurements of directly imaged brown dwarf GQ Lup B to search for exo-satellites}

\newcommand{\petit}{\texttt{petitRADTRANS}}
\newcommand{\emcee}{\texttt{emcee}}
\newcommand{\pmn}{\texttt{PyMultiNest}}
\newcommand{\moog}{\texttt{MOOG}}
\newcommand{\teff}{T$_{\rm eff}$}
\newcommand{\feh}{\rm{[Fe/H]}}
\newcommand{\vsini}{\ensuremath{v \sin i}\xspace}
\newcommand{\caltech}{Department of Astronomy, California Institute of Technology, Pasadena, CA 91125, USA}
\newcommand{\gps}{Division of Geological \& Planetary Sciences, California Institute of Technology, Pasadena, CA 91125, USA}
\newcommand{\ucsc}{Department of Astronomy \& Astrophysics, University of California, Santa Cruz, CA95064, USA}
\newcommand{\keck}{W. M. Keck Observatory, 65-1120 Mamalahoa Hwy, Kamuela, HI, USA}
\newcommand{\ucla}{Department of Physics \& Astronomy, 430 Portola Plaza, University of California, Los Angeles, CA 90095, USA}
\newcommand{\jpl}{Jet Propulsion Laboratory, California Institute of Technology, 4800 Oak Grove Dr.,Pasadena, CA 91109, USA}
\newcommand{\ucsd}{Department of Astronomy \& Astrophysics,  University of California, San Diego, La Jolla, CA 92093, USA}
\newcommand{\ucsdp}{Department of Physics, University of California, San Diego, La Jolla, CA 92093}
\newcommand{\ifahonolulu}{Institute for Astronomy, University of Hawai`i, 2680 Woodlawn Drive, Honolulu, HI 96822, USA}
\newcommand{\berkeley}{Department of Astronomy, University of California at Berkeley, CA 94720, USA}
\newcommand{\ames}{NASA Ames Research Centre, MS 245-3, Moffett Field, CA 94035, USA}
\newcommand{\northwestern}{Center for Interdisciplinary Exploration and Research in Astrophysics (CIERA), Northwestern University, 1800 Sherman, Evanston, IL, 60201, USA}
\newcommand{\UofA}{James C. Wyant College of Optical Sciences, University of Arizona, Meinel Building 1630 E. University Blvd., Tucson, AZ. 85721}

\correspondingauthor{Katelyn Horstman}
\email{khorstma@caltech.edu}

\author[0000-0001-9708-8667]{Katelyn Horstman}
\affiliation{\caltech}
\altaffiliation{NSF Graduate Research Fellow}

\author[0000-0003-2233-4821]{Jean-Baptiste Ruffio}
\affiliation{\ucsd}

\author[0000-0002-7094-7908]{Konstantin Batygin}
\affiliation{Division of Geological and Planetary Sciences California Institute of Technology, Pasadena, CA 91125, USA}

\author[0000-0002-8895-4735]{Dimitri Mawet}
\affiliation{\caltech}
\affiliation{\jpl}

\author{Ashley Baker}
\affiliation{\caltech}


\author[0000-0002-5370-7494]{Chih-Chun Hsu}
\affil{\northwestern}

\author[0000-0003-0774-6502]{Jason J. Wang (王劲飞)}
\affiliation{\northwestern}


\author[0000-0002-4361-8885]{Ji Wang (王吉)}
\affiliation{Department of Astronomy, The Ohio State University, 100 W 18th Ave, Columbus, OH 43210 USA}




\author[0000-0002-3199-2888]{Sarah Blunt}
\affiliation{\northwestern}

\author[0000-0002-6618-1137]{Jerry W. Xuan}
\affiliation{\caltech}

\author{Yinzi Xin}
\affiliation{\caltech}

\author[0000-0002-4934-3042]{Joshua Liberman}
\affiliation{\caltech, \UofA}

\author[0000-0003-2429-5811]{Shubh Agrawal}
\affiliation{\caltech}
\affiliation{Department of Physics and Astronomy, University of Pennsylvania, Philadelphia, PA 19104, USA}

\author[0000-0002-9936-6285]{Quinn M. Konopacky}
\affiliation{\ucsd}

\author{Geoffrey A. Blake}
\affiliation{Division of Geological and Planetary Sciences California Institute of Technology, Pasadena, CA 91125, USA}

\author[0000-0001-5173-2947]{Clarissa R. Do \'O}
\affiliation{\ucsdp}




\author{Randall Bartos}
\affiliation{\jpl}

\author{Charlotte Z. Bond}
\affiliation{UK Astronomy Technology Centre, Royal Observatory, Edinburgh EH9 3HJ, United Kingdom}

\author{Benjamin Calvin}
\affiliation{\caltech}
\affiliation{\ucla}

\author{Sylvain Cetre}
\affiliation{\keck}

\author[0000-0001-8953-1008]{Jacques-Robert Delorme}
\affiliation{\keck}
\affiliation{\caltech}

\author{Greg Doppmann}
\affiliation{\keck}

\author[0000-0002-1583-2040]{Daniel Echeverri}
\affiliation{\caltech}

\author[0000-0002-1392-0768]{Luke Finnerty}
\affiliation{\ucla}

\author[0000-0002-0176-8973]{Michael P. Fitzgerald}
\affiliation{\ucla}

\author[0000-0001-5213-6207]{Nemanja Jovanovic}
\affiliation{\caltech}

\author{Ronald L\'opez}
\affiliation{\ucla}

\author[0000-0002-0618-5128]{Emily C. Martin}
\affiliation{\ucsc}

\author{Evan Morris}
\affiliation{\ucsc}

\author{Jacklyn Pezzato}
\affiliation{\caltech}

\author[0000-0003-4769-1665]{Garreth Ruane}
\affiliation{\caltech}
\affiliation{\jpl}

\author[0000-0003-1399-3593]{Ben Sappey}
\affiliation{\ucsd}

\author{Tobias Schofield}
\affiliation{\caltech}

\author{Andrew Skemer}
\affiliation{\ucsc}

\author{Taylor Venenciano}
\affiliation{Physics and Astronomy Department, Pomona College, 333 N. College Way, Claremont, CA 91711, USA}

\author[0000-0001-5299-6899]{J. Kent Wallace}
\affiliation{\jpl}

\author[0000-0003-0354-0187]{Nicole L. Wallack}
\affiliation{Earth and Planets Laboratory, Carnegie Institution for Science, Washington, DC 20015, USA}

\author{Peter Wizinowich}
\affiliation{\keck}



\begin{abstract}
  GQ Lup B is one of the few substellar companions with a detected cicumplanetary disk, or CPD. Observations of the CPD suggest the presence of a cavity, possibly formed by an exo-satellite. Using the Keck Planet Imager and Characterizer (KPIC), a high contrast imaging suite that feeds a high resolution spectrograph (1.9-2.5 microns, R$\sim$35,000), we present the first dedicated radial velocity (RV) observations around a high-contrast, directly imaged substellar companion, GQ Lup B, to search for exo-satellites. Over 11 epochs, we find a best and median RV error of 400-1000 m/s, most likely limited by systematic fringing in the spectra due to transmissive optics within KPIC. {With this RV precision, KPIC is sensitive to exomoons 0.6-2.8\% the mass of GQ Lup B ( \(\sim 30 \, M_{\text{Jup}}\)) at separations between the Roche limit and \(65 \, R_{\text{Jup}}\), or the extent of the cavity inferred within the CPD detected around GQ Lup B. Using simulations of HISPEC, a high resolution infrared spectrograph planned to debut at W.M. Keck Observatory in 2026, we estimate future exomoon sensitivity to increase by over an order of magnitude, providing sensitivity to less massive satellites potentially formed within the CPD itself. Additionally, we run simulations to estimate the amount of material that different masses of satellites could clear in a CPD to create the observed cavity. We find satellite-to-planet mass ratios of $q > 2 \times 10^{-4}$ can create observable cavities and report a maximum cavity size of \(\sim 51 \, R_{\text{Jup}}\) carved from a satellite.}
 
\end{abstract}



\section{Introduction}
\label{sec:introduction}

\subsection{Exo-satellites as potential planet formation probes}
Understanding the striking diversity of planetary systems requires an exploration of how they form and evolve, yet fundamental questions remain unanswered, such as how common our own solar system is. By examining the over 200 moons in our solar system, ranging over a variety of inclinations, eccentricities, and compositions, we have gained a wealth of knowledge about how our solar system formed and how it achieved its current architecture. Similarly, we can also study exomoons, or moons beyond our solar system, to gain valuable insights into how planets form, both within and beyond our solar system \citep{HellerReview2014, Teachey2024}.

The satellites around exoplanets, or exo-satellites/ exomoons, have been predicted to form in a variety of ways: in the circumplanetary disk (CPD) surrounding an exoplanet; from capture, such as Neptune's moon Triton \citep{Agnor2006}; collisions with protoplanets, such the Moon \citep{Canup2001}; or even disk instabilities, such as brown dwarf binaries \citep{Lazzoni2020}. When forming exo-satellites from the dust and gas surrounding the companion, the CPD is believed to be a decretion disk where dust of critical size becomes trapped within the disk and fragments into satellitesimals once the disk reaches a supersolar metallicty. These satellitesimals can then grow and migrate inward due to gas drag \citep{Batygin2020}, implying moons are a natural consequence of planet formation. The typical CPD total dust mass relative to the planet is around $10^{-4}$ \citep{Canup2006, Sasaki2010} which is consistent with the mass ratio of the Galilean satellites around Jupiter. Additionally, this measurement of dust mass is also consistent with observations of the CPD around PDS 70 c from ALMA continuum observations \citep{Benisty2021}. 

Theoretical models also suggest that it is possible massive planets form even more massive moons following the scaling $\mathrm{m\propto M^{3/2}}$, with $m$ and $M$ the masses of the moon and the planet respectively (based on equation 43 in \cite{Batygin2020}). Therefore, satellites around brown dwarfs could likely reach the critical mass (\(\sim 10 M_{\Earth}\) from \citet{Piso2014}) necessary for runaway gas accretion, leading to much higher mass ratios than the Galilean moons (\(> 10^{-3}\)). Recent simulations suggest that it is possible to form a single, massive moon in the CPD rather than  smaller mass moons. \citet{Fujii2020} found there are favorable viscous parameters that lead to the formation and stability of singular moons around a gas giant planets. Additionally, 3D hydrodynamical simulations propose eccentric companions can incite retrograde CPDs capable of forming retrograde satellites, suggesting formation pathways for retrograde, higher mass satellites other than capture \citep{Chen2022}.

\subsection{Exo-satellite candidates and observations} 
The field of exosolar satellite theory and detection has experienced significant growth over the past decade and is being used to answer questions about planet formation, within and beyond our own solar system. Transiting surveys have been the dominant method used to conduct exomoon searches and have yielded two current, yet contested, exomoon candidates \citep{Teachey2018a, Kipping2022, Heller2024, Kipping2024}. Studies continue to look for exo-satellites using transits to place limits on the detectable masses and probe different system architectures, even when no strong evidence of a satellite is present \citep{Limbach2021, Ohno2022, Ehrenreich2023}. Other exo-satellite detection techniques include microlensing, which identified candidates in the MOA-2011-BLG-262 and MOA-2015-BLG-337 systems \citep{Bennett2014, Miyazaki2018}, searching for sodium and potassium due to geological activity on volcanic worlds \citep{Oza2019, Gebek2020}, and looking for exoplanet-satellite interactions in radio wavelengths \citep{Pineda2017, Narang2023, Kao2024}.

Due to their larger size and therefore larger Hill Spheres, directly-imaged exoplanets may be more likely to harbor larger satellites. \citet{Lazzoni2020} claims high contrast imaging detected an exomoon candidate (estimated \(1 \, M_{\text{Jup}}\) orbiting a directly imaged BD, DH Tau B). Additionally, the combination of high contrast imaging and high resolution spectroscopy is already a viable method to detect small satellites around directly-imaged planets. Motivated by \citet{Teachey2018a}, \citet{Vanderburg2018} started exploring the detectability of exo-satellites around directly imaged planets using Doppler spectroscopy. Shortly after, \citet{Vanderburg2021} placed the first limits on binary planets and exo-satellites using OH-Suppressing Infrared Integral Field Spectrograph (OSIRIS) radial velocity observations of HR 8799 b, c, and d from \citet{Ruffio2021} and found sensitivity to companions more massive than \(2 \, M_{\text{Jup}}\), or with a mass ratio greater than 20\%, for orbital periods less than five days. Influenced by this work, the Keck Planet Imager and Characterizer (KPIC) team analyzed existing observations of the BD companion HR 7672 B and achieved 2 km/s RV precision for 5 min exposures, demonstrating sensitivity to moon-to-planet mass ratios between 1 and 4\% \citep{Ruffio2023}.

Although there have been quite a few notable exomoon candidates, there has still not been a definitive detection. So, it is natural to wonder where the best places to look for exo-satellites are. One option beginning to be explored is within the CPD around a still forming exoplanet, especially with the first CPD detections around companions PDS 70 c and GQ Lupi B \citep{Benisty2021, Stolker2021}.

\subsection{The curious case of GQ Lupi B}
GQ Lupi B was discovered by \citet{Neuh2005} at a separation of approximately 100 AU from host star GQ Lupi in the Lupus star forming region. With an apparent magnitude of $13.1 \pm 0.1$ and a contrast ratio of $4 \times 10^{-3}$ in K-band \citep{Neuh2005}, it is a favorable companion to observe in the near-infrared. The companion historically had an uncertain mass between \(10-40 \, M_{\text{Jup}}\) \citep{Marois2007, McElwain2007, Seifahrt2007} for its measured temperature of $2650 \pm 100 \, \text{K}$ \citep{Seifahrt2007}, but recent studies have constrained the mass to \(30 \, M_{\text{Jup}}\) \citep{Stolker2021, Xuan2024}.
 
Early observations of GQ Lup B, based on HST and Subaru data, suggest that the system is still in its formation phase and is actively accreting based on detections of H$\alpha$ emission \citep{Marois2007}. Further follow up has confirmed the accretion signature in H$\alpha$ \citep{Zhou2014, Wu2017, Stolker2021} and Pa$\beta$ \citep{Seifahrt2007, Demars2023}. GQ Lup B is estimated to be accreting at a rate of \(M \approx 10^{-6.5} M_J \, \text{yr}^{-1}\) \citep{Stolker2021} while its emission line variability suggests magnetospheric accretion \citep{Demars2023}. GQ Lup B is also measured to have a $v \sin i$ of 5-6 km/s using VLT/CRIRES and KPIC data respectively, indicating that it is a slow rotator \citep{Schwarz2016, Xuan2024}. It is thought to be a slow rotator since it is still young (2-5 Myr) and is expected to spin up as it continues to accrete more material and contract \citep{Donati2012}.

The inclination of the disk around GQ Lupi is reported to be $60.5 \pm 0.5$ degrees with respect to the line of sight, and it is observed to be misaligned with the spin axis of the star by approximately $30$ degrees \citep{MacGregor2017, Wu2017}. Observations by \citet{vanHolstein2021} show spiral-like structure in the circumstellar disk which may be due to gravitational interaction with GQ Lup B. Radial velocity measurements of $2.0 \pm 0.4 \ \mathrm{km/s}$, along with astrometry data from VLT/NACO and HST, have been utilized to constrain the orbit of GQ Lup B \citep{Schwarz2016}. Its orbit exhibits a mutual inclination of $84 \pm 9$ degrees relative to the circumstellar disk \citep{Stolker2021}. The orientation of GQ Lup B's orbit relative to the circumstellar disk, combined with measurements of CO, H$_2$O, and metallicity, suggests that GQ Lup B formed via gravitational collapse \citep{Stolker2021, Demars2023, Xuan2024}. Adding to its interesting system architecture, a second companion, GQ Lup C, has been detected at an approximate separation of 2400 AU \citep{Alcala2020A, Lazzoni2020C}. 

Interestingly, a potential protolunar disk has been detected from infrared excess around GQ Lup B. Using ALMA observations of the GQ Lup system, \citet{MacGregor2017} placed an upper limit of \(0.04 
~M_{\text{Earth}}\) on the dust mass in the CPD surrounding GQ Lup B, arguing that a non-detection of the disk could be due to its compact and optically thick nature. However, a recent detection of the CPD around GQ Lup B by \citet{Stolker2021} suggests that the non-detection of dust grains by ALMA may actually be due to the depletion of dust in the inner disk due to satellite formation. \citet{Stolker2021} claims the existence of a \(65 \pm 1 \, R_{\text{J}}\) cavity in the CPD might be caused by one or multiple young moons carving away material, since the size of the cavity is larger than the expected dust sublimation radius calculated from the effective temperature of GQ Lup B. While this paper was in preparation, additional JWST observations of GQ Lup B also investigated the cavity within the CPD, finding a cavity size ranging between \(8-40 \, R_{\text{Jup}}\) depending on the model CPD proposed \citep{Cugno2024}.

Looking for exoplanets carving the gaps and cavities in protoplanetary disks is already a common strategy used to search for new exoplanets and is therefore an appealing way to search for a moon in the cavity of the CPD around GQ Lup B \citep{Ruane2017}. Both ALMA and SPHERE data have revealed dust substructures and cavities within protoplanetary disks that could potentially be attributed to the presence of planets \citep{Feng2018, Torres2021} and notably two planets were discovered within the gap of the disk in the PDS 70 system, illustrating the effectiveness of this method \citep{Keppler2018, Haffert2019}. 

GQ Lup B is the ideal candidate to probe for satellites because of the detected cavity in the CPD, possibly caused by moon formation, the likelihood of this young brown dwarf to harbor proportionally more massive moons, and the ability of KPIC to achieve high precision RV measurements via direct imaging spectroscopy.

\subsection{Outline}
In this paper, we aim to search for an exo-satellite in the cavity of the CPD surrounding brown dwarf GQ Lup B. In \autoref{sec:KPIC observations} and \autoref{sec: Data Reduction}, we present the first dedicated RV survey to search for satellites around a high contrast companion using KPIC and explain how we obtained our RV measurements. In \autoref{sec:exomoon sensitivity} we present the exo-satellite sensitivity achieved with KPIC, and compare the results obtained to the predicted cavity carving nature of exomoons in \autoref{sec:carving}. Finally, in \autoref{sec:Discussion} we conclude by discussing the future prospects for exo-satellite searches with high resolution spectrographs and hypothesize about their occurrence rates.

\begin{table*}
  \centering
\begin{tabular}{|c|c|c|c|c|} 
 \hline
  Date (UT) & Number of Exposures & Exposure time (sec) & Seeing & Throughput \\ 
 \hline
2021-04-24 & 6 & 600 &$1.0^{\prime\prime}$& 1.6\%  \\
2022-07-20 & 39 & 60 &$0.6^{\prime\prime}$& 2.0\% \\
2022-07-20 & 6 & 600 &$0.6^{\prime\prime}$& 2.0\% \\
2022-07-21 & 4 & 600 &$0.5^{\prime\prime}$& 3.0\% \\
2022-07-23 & 2 & 600 &$0.6^{\prime\prime}$& 2.2\% \\
2023-05-06 & 21 &180 &$0.7^{\prime\prime}$& 1.7\% \\
2023-06-04 & 18 &180 &$0.6^{\prime\prime}$ & 0.8\% \\
2023-06-21 & 34 &180 & $0.7^{\prime\prime}$ & 1.8\% \\
2023-06-23 & 33 &180 & $0.4^{\prime\prime}$ & 2.8\% \\
2023-06-24 & 32 &180 & $0.9^{\prime\prime}$ & 2.2\% \\
2023-06-29 & 32 &180 & $0.5^{\prime\prime}$ & 3.8\% \\
2023-07-02 & 25 &180 & $0.6^{\prime\prime}$ & 2.4\% \\
 \hline
\end{tabular}
\caption{$K$-band observations of GQ Lup B with KPIC. The end-to-end throughput is measured from top of the atmosphere and is a better metric of performance than seeing for KPIC. We report the $95\%$ percentile throughput over the $K$ band, averaged over all frames.}
\label{tab:observations}
\end{table*}

\section{KPIC Observations} 
\label{sec:KPIC observations}
We carried out 11 observations of GQ Lup B with KPIC ($R\sim35,000$) in K-band (1.94-2.49~$\mu$m) \citep{Mawet2017,Delorme2021b}, a high contrast imaging suite that feeds the high resolution spectrograph NIRSPEC \citep{McLean1998, Martin2018}, between 2021 and 2023. The first 4 observations were completed during KPIC Phase 1 (2019-2021) \citep{Delorme2021b}. Since 2022, KPIC received several upgrades \citep{Echeverri2022}, and in May 2023, we began the first dedicated RV exo-satellite survey to monitor GQ Lup B. Since GQ Lup B is at a relatively low elevation in the northern hemisphere (DEC = \SI{-35}{\degree} \SI{39}{\arcminute} \SI{05.0539}{\arcsecond}), we observed the brown dwarf for only 1.5-2.5 hours per night for 7 nights between May and July 2023. A summary of our observations is provided in \autoref{tab:observations}. 

We use the same observing techniques as in \citet{Wang2021}, except we switch between the two highest performing fibers, determined by measuring which of the four fibers has the best end-to-end throughput, to aid in background subtraction between exposures. The relative astrometry of GQ Lup B was computed using \href{http://whereistheplanet.com/}{whereistheplanet.com} \citep{whereistheplanet} so the position of the fiber is correctly aligned with the companion on a given night. 

\section{Data Reduction}
\label{sec: Data Reduction}
Spectra were reduced using the KPIC Data Reduction Pipeline (DRP) \footnote{\url{https://github.com/kpicteam/kpic_pipeline}} following the same procedure as described in \citet{Wang2021}. In summary, the KPIC DRP performs background subtraction, bad pixel correction, and spectral trace calibration to determine the location and width of each of the nine NIRSPEC spectroscopic orders, orders 31-39, on the detector for each of KPIC's four fibers. Using calibration data taken during the night of observation, the spectrum of an early M giant star is used to derive a wavelength solution for each spectroscopic order. Spectral lines from the M-calibrator star, in this case primarily HIP 81497, and telluric lines from the atmosphere are modeled with a PHOENIX model \citep{Husser2013} and the Planetary Spectrum Generator \citep{Villanueva2018} respectively to obtain best fit parameters for the final wavelength solution in each order. Additionally, during our observing sequence on GQ Lup B, we take intermittent observations of the host star, GQ Lup, to account for additional light from the host star leaking into the science fiber model.

\subsection{Forward model and likelihood}
\label{sec:FM}

\begin{figure*}
  \centering
  \includegraphics[trim={0cm 0cm 0cm 0cm},clip,width=1\linewidth]{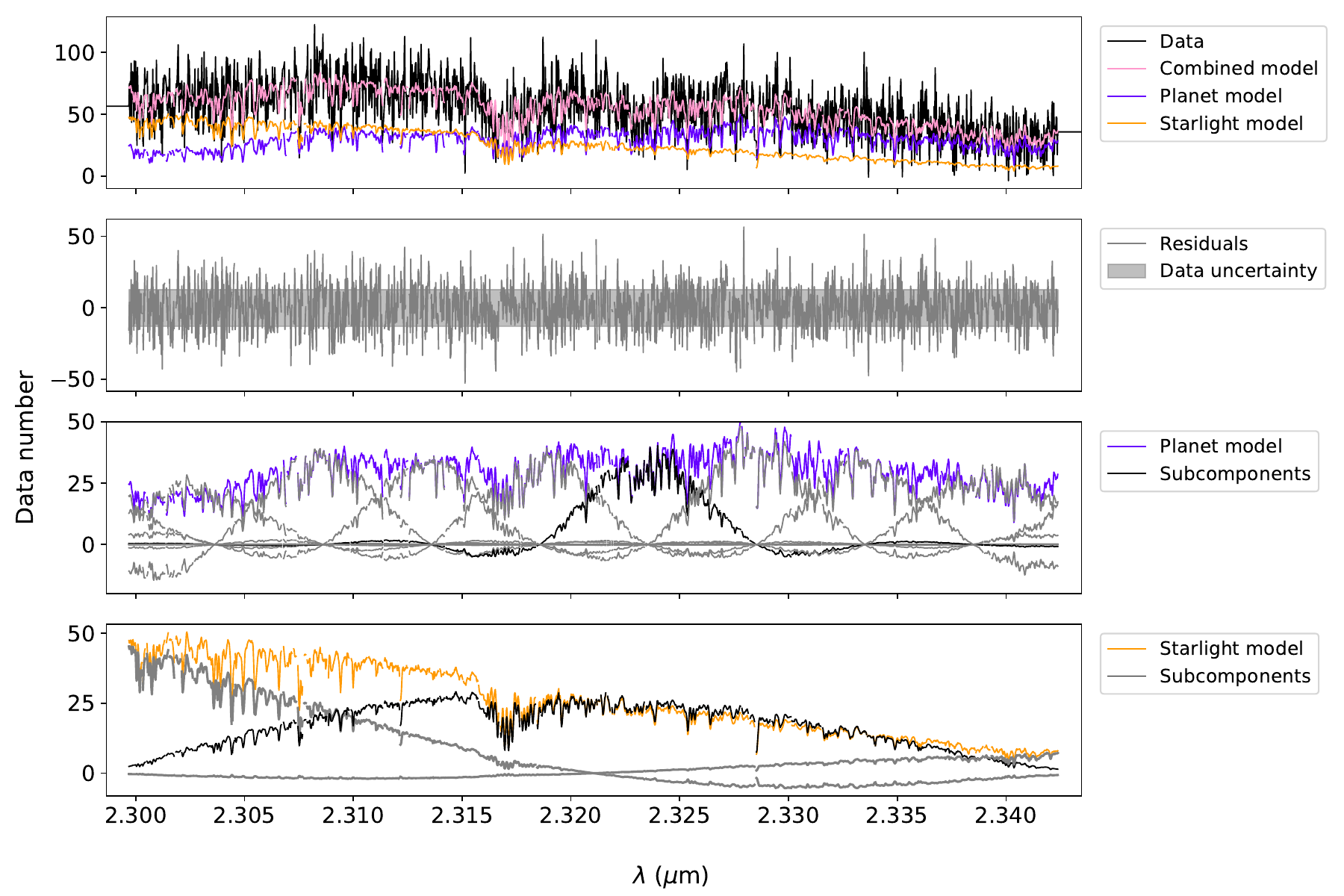}
  \caption{An example GQ Lup B spectrum and forward model from a single exposure on 06/29/2023 (UT). \textbf{Top:} The full spectrum is in black, while the combined model, consisting of the planet model (purple) and the stellar model (orange), is in pink. \textbf{Second panel:} The residuals (data-model) are shown in gray, while the shaded gray region represents the data uncertainty measured by the KPIC DRP. The data uncertainties are derived from optimal extraction \citep{Horne1986}, which fits a vertical Gaussian profile to the horizontal spectral trace on the detector at each wavelength, outlined in Section 3.1 of \citet{Wang2021}. \textbf{Third panel:} The forward model of the companion spectrum taken from a BT-Settl atmospheric model ($\log g=4.0$; $T_{\mathrm{eff}}=2700\,\mathrm{K}$, \citet{Allard2012}). The ten spline nodes used to model the continuum are shown in gray and black.} \textbf{Bottom:} The forward model of the starlight derived from empirical observations of GQ Lup to account for speckle light leaking into the fiber. The three spline nodes used to model the continuum are shown in gray and black.
\label{fig:fm}
\end{figure*}

We jointly model the host and companion spectra by using the python package \texttt{breads}\footnote{\url{https://breads.readthedocs.io/en/latest/}} \citep{Agrawal2023} 
to measure the RV of GQ Lup B for each exposure following the same method used in \citet{Ruffio2023}. 

We define our forward model as,
\begin{equation}
     {\bm d} = {\bm M}_{\mathrm{RV}}{\bm \phi} + {\bm n},
\label{eq:fm_eq}
\end{equation}
where ${\bm d}$ is the data vector of size $N_d$, ${\bm M}_{\mathrm{RV}}$ is the linear model, ${\bm \phi}$ are the linear parameters, and ${\bm n}$ is a random vector of the noise with a diagonal covariance matrix ${\bm \Sigma}$, where ${\bm \Sigma}={\bm \Sigma}_0 s^2$. ${\bm \Sigma}_0$ is defined using both the data vector and the standard deviation of the noise, and is multiplied by a free parameter scaling factor $s^2$ to account for any underestimation of the noise.

Observations of GQ Lup, the host star, are used to empirically derive both a stellar spectrum and instrument transmission. The stellar spectrum is necessary to model stellar speckles leaking into the fiber placed on the position of the companion. The companion is modeled using a BT-Settl atmospheric model\footnote{\url{https://phoenix.ens-lyon.fr/Grids/BT-Settl/CIFIST2011c/}} ($\log g=4.0$; $T_{\mathrm{eff}}=2700\,\mathrm{K}$, \citealp{Allard2012}) broadened by the empirical line spread function (LSF), then multiplied by the telluric spectrum and instrument transmission profile. To model the stellar and planet continuum, we control the number of nodes used in a $3^{\rm rd}$ order spline model. To account for inaccuracies in the continuum of the atmospheric model, ten spline nodes are used in each spectral order ($\Delta \lambda \sim0.05\,\mu$m) for the planet model. This is equivalent to a 200 pixel wide high-pass filter, to balance the number of parameters modeled with the optimal high-pass filter scale of 100 pixels found in \citet{Xuan2022}. Three spline modes are used to model the speckle continuum to account for speckles appearing in the fiber location as function of wavelength. The RV of GQ Lup B is the only non-linear parameter we fit for in our forward modeling framework. An example of our forward model for a single exposure is shown in \autoref{fig:fm}.

KPIC residuals show systematic fringing, or periodic oscillations in the continuum flux as a function of wavelength, due to Fabry-P\'{e}rot etalons created by the transmissive optics in both NIRSPEC \citep{Hsu2021} and KPIC \citep{Finnerty2022}. Fringing is a persistent problem and can greatly affect our ability to fit the data. For high signal-to-noise (S/N) observations, the fringing amplitude can reach up to 15\% of the stellar continuum \citep{Finnerty2022}. Several different attempts have been made to mitigate the fringing signal. For temporal observations of Hot Jupiters, \citet{Finnerty2023} removed the time-varying fringing signal attributed to the KPIC optics by using PCA analysis. Additionally, \citet{Xuan2024AB} incorporated a physical fringing model fit to the contaminated residuals of their spectra to account for the extra systematics. After exploring various fringing mitigation options, we follow the same procedure as in \cite{Ruffio2023} and apply a Fourier filter to remove the main frequencies associated with the periodic fringing signal. 

The likelihood function is defined from a multivariate Gaussian distribution as,
\begin{multline}
    \mathcal{L}(\mathrm{RV},{\bm \phi},s^2) =\frac{1}{\sqrt{(2\pi)^{N_d}\vert{\bm \Sigma}_0\vert s^{2N_d}}}\\  \exp\left[-\frac{1}{2s^2}({\bm d}-{\bm M}_{\mathrm{RV}}{\bm \phi})^{\top}{\bm \Sigma}_0^{-1}({\bm d}-{\bm M}_{\mathrm{RV}}{\bm \phi})\right].
\end{multline} 
We find the maximum likelihood RV for each exposure using a linear least square solver on a grid of RV values ranging from $-400$ to $400\,$km/s in steps of $0.2\,$km/s. We derive $1\sigma$ RV uncertainties according to Equation 10 in \citet{Ruffio2021} from each RV posterior distribution. Each spectral order is fit separately to the determine the RV. Only the three orders with both sufficient stellar and telluric lines in K-band were used in this analysis: $2.29-2.34\,\mu$m (order 33), which coincides with the CO bandhead, $2.36-2.41\,\mu$m (order 32), and $2.44-2.49\,\mu$m (order 31), giving three RV measurements per exposure, as shown in \autoref{fig:rvs}.

\subsection{RV measurements}

\begin{figure*}
  \centering
  \includegraphics[trim={0cm 0cm 0cm 0cm},clip,width=1\linewidth]{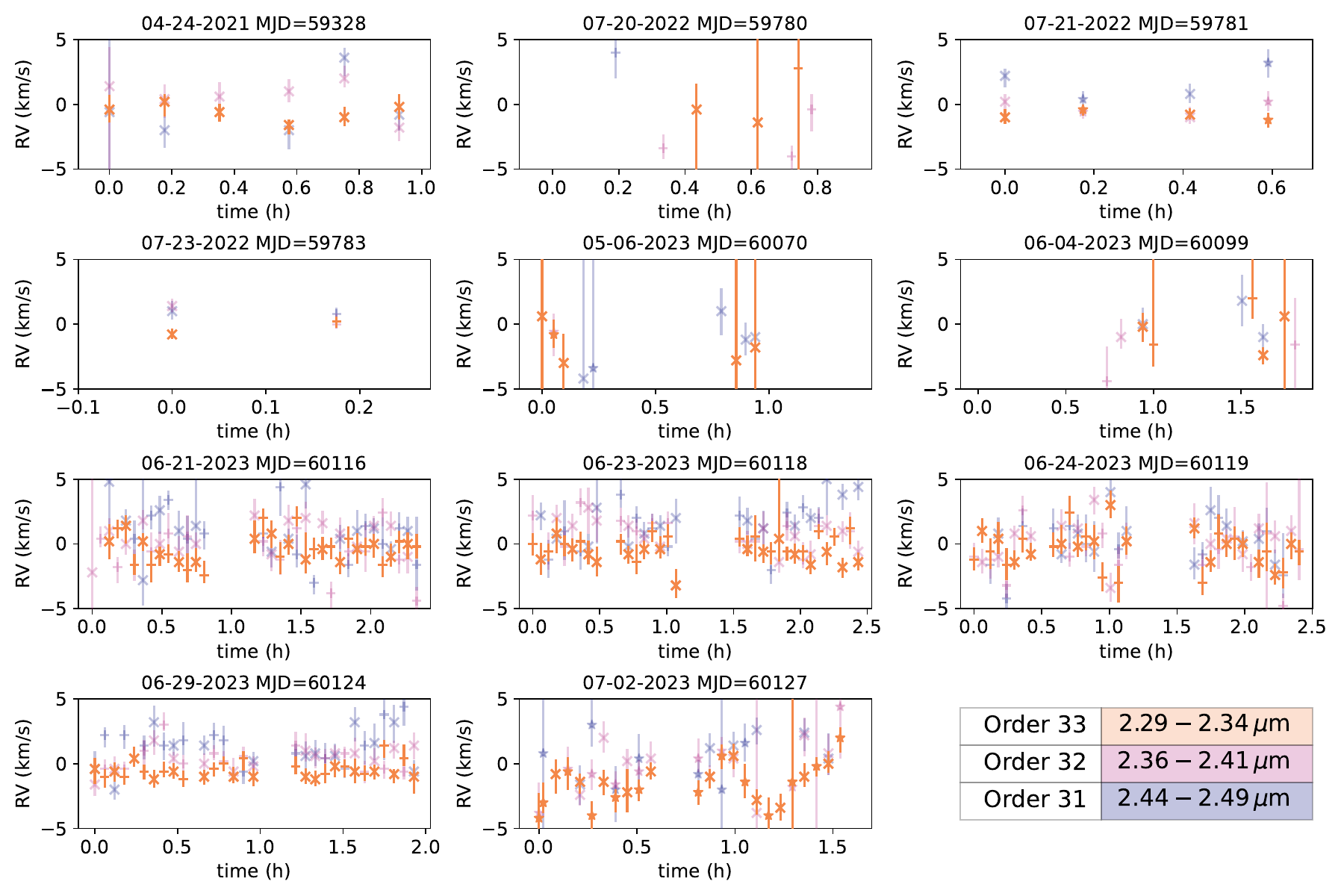}
  \caption{Measured RVs of GQ Lup B with KPIC. Each panel represents a different epoch. In the forward modeling framework used, each spectral order is fit separately to measure a RV. Orange represents order 33, pink represents order 32, and purple represents order 31. Each marker shape denotes a specific KPIC fiber where 'x' represents fiber 2, '+' represents fiber 3, '$\star$' represents fiber 4. We limit our final data set to RVs with a CCF S/N \textgreater 3 and a measured value between -5 to 5 km/s.}
  \label{fig:rvs}
\end{figure*}

\begin{table*}
    \centering
\begin{tabular}{|c|ccc|ccc|ccc|}
\hline
MJD & Order 33 RV& Error & Inflated Err & Order 32 RV & Error& Inflated Err& Order 31 RV & Error& Inflated Err\\
 & (km/s) & (km/s) & (km/s) & (km/s) & (km/s) &(km/s) & (km/s) & (km/s) & (km/s) \\
\hline
59328 & -0.78 &   0.32 &    -    &  0.69 &  0.45 &    0.62 & -0.16 &   0.59 &    1.39 \\
59780 & -0.25 &  17.82 &    -    & -3.11 &  0.61 &    1.29 &  4.00 &  47.86 &    -    \\
59781 & -0.77 &   0.24 &    -    & -0.37 &  0.33 &    0.35 &  1.37 &   0.39 &    0.71 \\
59783 & -0.33 &   0.29 &    -    &  0.81 &  0.39 &    -    &  0.87 &   0.43 &    -    \\
60070 & -1.24 &   0.99 &    -    & -0.60 &  1.65 &    -    & -0.48 &   1.04 &    -    \\
60099 & -1.82 &   0.59 &    -    & -1.62 &  1.04 &    1.30 & -0.22 &   0.75 &    0.94 \\
60116 & -0.34 &   0.16 &    0.19 & -0.04 &  0.18 &    0.28 &  0.71 &   0.22 &    0.35 \\
60118 & -0.37 &   0.14 &    0.15 &  0.54 &  0.18 &    0.25 &  1.44 &   0.22 &    0.29 \\
60119 & -0.44 &   0.18 &    0.25 & -0.20 &  0.19 &    0.31 &  0.15 &   0.26 &    0.38 \\
60124 & -0.57 &   0.12 &    -    &  0.35 &  0.14 &    0.17 &  0.94 &   0.19 &    0.26 \\
60127 & -1.21 &   0.20 &    0.30 & -0.39 &  0.28 &    0.32 &  0.98 &   0.42 &    0.45 \\
\hline
\end{tabular}
\label{tab:av_rv}
\caption{Combined nightly RV of GQ Lup B for each order with KPIC. For each order, irrespective of fiber number, the mean weighted RV, $1\sigma$ error, and inflated error are reported. The inflated errors normalize the reduced chi-squared, $\chi^2_{r}$, value to unity to compensate for underestimated systematic errors within a single epoch. If $\chi^2_{r}<1$, we report the $1\sigma$ error only and leave the inflated error column blank. Since there are inconsistencies between combined RVs in a single epoch for orders 31-33, we assume a different zero point RV for each order and fiber pair with \texttt{RVSearch} to jointly estimate the offsets between them. A full table with measured RVs and errors for each exposure, as seen in \autoref{fig:rvs}, can be found attached to this table.}
\end{table*}

Following the procedure outlined in the previous section, we measure the barycentric-corrected RV for each order, shown in \autoref{fig:rvs}. We do not correct for the orbital motion of GQ Lup B, as over the timescale of our observations, its radial velocity is expected to change by less than $10^{-12}$ km/s due to its wide separation. For each exposure, we compute a cross-correlation function (CCF) S/N to estimate the companion flux as a function of RV as in \citet{Ruffio2019, Wang2021}. To prevent spurious detections, we limit our final data set to RVs with a CCF S/N \textgreater 3 and a measured value between -5 to 5 km/s. 

Over 11 epochs we use in this analysis, we find a best and median RV error of 400 m/s and 1 km/s respectively for individual exposures, most likely limited by systematic fringing instead of photon noise.

\section{Exo-satellite sensitivity around GQ Lup B with KPIC}
\label{sec:exomoon sensitivity}

We use the open-source Python package \texttt{RVSearch}\footnote{\url{https://github.com/California-Planet-Search/rvsearch}} \citep{Rosenthal2021} to look for possible satellites around GQ Lup B and derive the sensitivity of our KPIC RV time series. \texttt{RVSearch} is a planet search algorithm developed by the California Legacy Survey to search for periodic signals in an RV time series \citep{Howard2016,Rosenthal2021,Fulton2021}. The algorithm searches over orbital periods defined by the user, computing a the difference in Bayesian Information Criterion (BIC) between a model including a companion signal and a model with no companion signal \citep{Rosenthal2021}. For each iterative search, the algorithm fits the histogram of $\Delta$BIC periodogram values using the power law noise model described in \cite{Howard2016} to derive a detection threshold for orbiting companions.

\begin{figure*}
\gridline{\fig{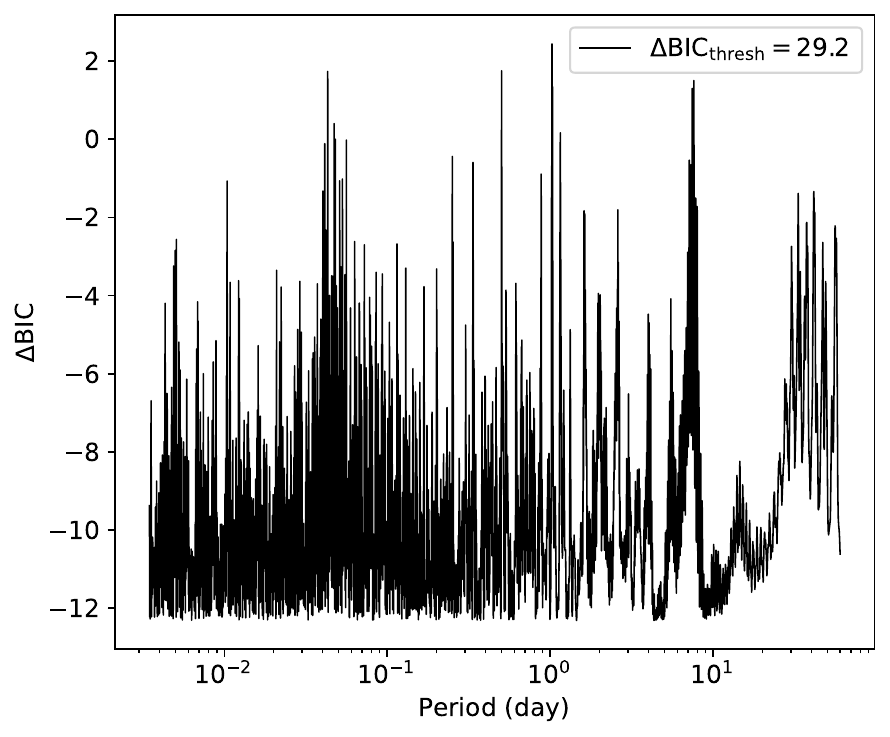}{0.49\textwidth}{(a) Periodogram}
		\fig{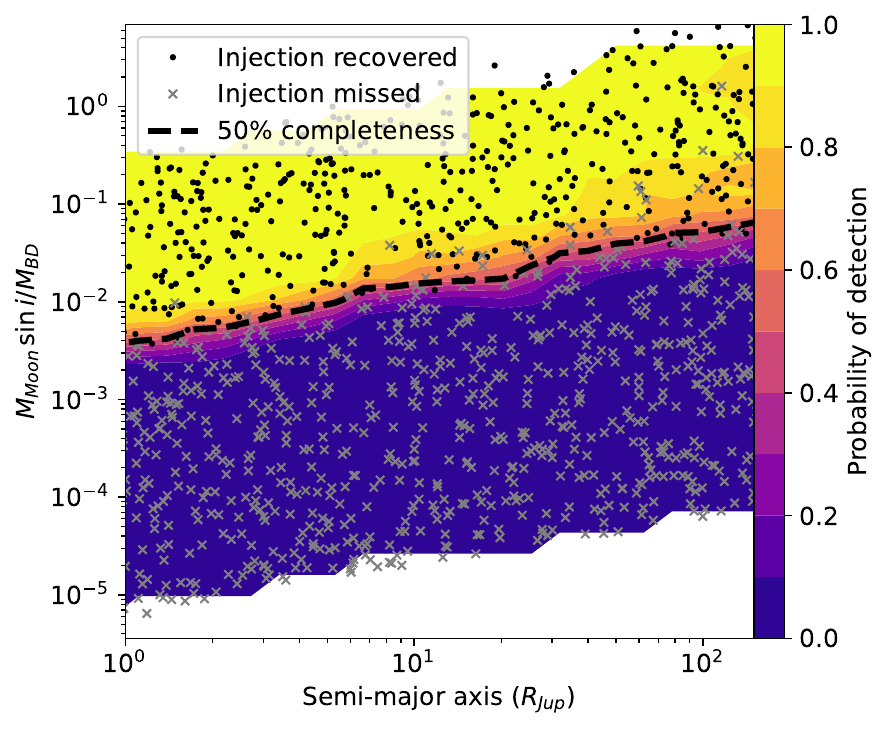}{0.49\textwidth}{(b) Completeness}
          }
\caption{Exo-satellite detection limits around GQ Lup B with the Keck Planet Imager and Characterizer (KPIC) using the open-source python module \texttt{RVsearch} \citep{Rosenthal2021}. \textbf{Left:} Periodogram of the RV times series shown in \autoref{fig:rvs} expressed a $\Delta$BIC comparing a model with and a model without a satellite signal. The empirical detection threshold is indicated in the legend and is much higher than the power expressed for the strongest periodic signal, indicating the model without a satellite signal is favored over a model with a satellite signal. \textbf{Right:} Exo-satellite completeness derived from injection and recovery tests. Recovered injections are standard circles, while missed injections are crosses. The 50\% completeness contour represents the mass ratio and orbital period space where the probability of satellite detection is at least 50\% based on injection and recovery tests.}
\label{fig:rvsearch_GQLupB}
\end{figure*}

\begin{figure*}
  \centering
  \includegraphics[trim={0cm 0cm 0cm 0cm},clip,width=1\linewidth]{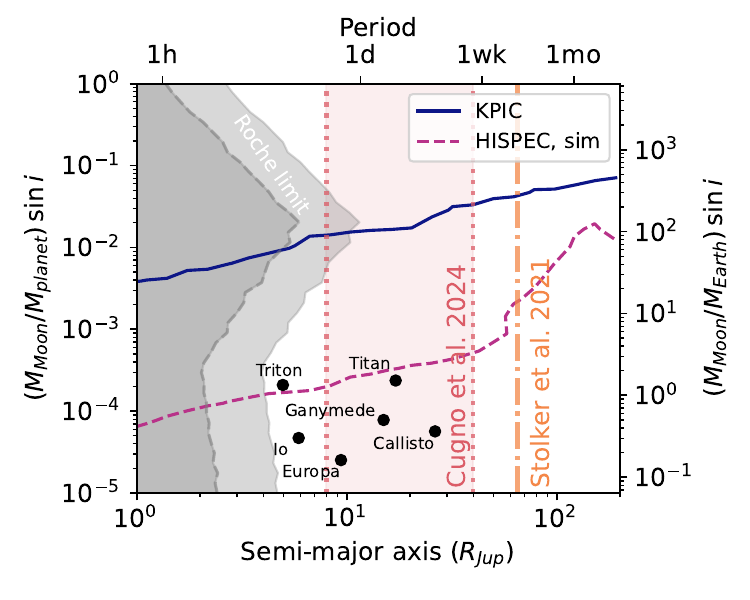}
  \caption{Exo-satellite sensitivity of KPIC around GQ Lup B. The current sensitivity of KPIC given the RV time series in \autoref{fig:rvs} is shown as blue line. KPIC is sensitive to exo-satellites 2.8\% the mass of GQ Lup B (\(30 \, M_{\text{Jup}}\)) at a separation of \(65 \, R_{\text{Jup}}\) and between 0.6-1.0\% at the intersection of the rigid and fluid Roche limits respectively. The expected sensitivity of a next generation high resolution spectrograph, HISPEC, is shown as a dashed purple line. To create the simulated RV time series of GQ Lup B using HISPEC, we assume 18 hours of observation spread across 6 nights and that each 180s exposure has an average RV sensitivity of 22.8 m/s. The darker and lighter gray dashed lines represent the Roche limit for rigid and fluid satellites respectively. The orange dashed-dotted line represents the measured radius of the CPD by \citet{Stolker2021}, \(65 \, R_{\text{Jup}}\), while the dotted pink lines represent the range of model dependent CPD radii measured from JWST data by \citet{Cugno2024}, \(8-40 \, R_{\text{Jup}}\). Several solar system moons are plotted for reference.}
  \label{fig:recovery}
\end{figure*}

\autoref{fig:recovery} shows the exo-satellite detection probability around GQ Lup B. Due to the different properties of each KPIC fiber, we assume a different zero point RV for each fiber and use \texttt{RVSearch} to jointly estimate the offsets between them. The same approach typically used for combining datasets from different RV instruments. We also use this same feature to account for possible systematics, primarily in the wavelength solution, between different NIRSPEC orders, as shown in \autoref{tab:av_rv}. 

If a satellite was indeed forming within the cavity, it would have an orbit of less than two weeks, similar to the Galilean moons. \autoref{fig:rvsearch_GQLupB} shows that the strongest periodic signal in the RV time series falls below the threshold necessary to claim a satellite detection. Additionally, $\Delta$BIC$\leq0$ for the majority of periods sampled, confirming the model including a satellite signal is never favored over the model without a satellite signal.

Although no exo-satellites were detected, we use \texttt{RVSearch} to perform injection-recovery tests to determine what moons KPIC could detect given its current RV precision. \autoref{fig:recovery} shows KPIC's exo-satellite sensitivity within the measured cavity of GQ Lup B, given a GQ Lup B mass of \(30 \, M_{\text{Jup}}\) \citep{Stolker2021, Xuan2024}. The 50\% completeness contour, shown in the right panel of \autoref{fig:rvsearch_GQLupB} and in \autoref{fig:recovery}, shows the mass ratios and orbital periods of exomoons KPIC has the probability of recovering 50\% of the time. Using injection-recovery tests, \texttt{RVSearch} creates a simulated exomoon signal, as described in \citep{Rosenthal2021}, with a period and $M \sin{i}$ from log-uniform distributions, and eccentricity from an empirically calibrated beta distribution \citep{Kipping2013} and injects it into the existing RV time series data. If the signal surpasses the $\Delta$BIC threshold, finding that a model including a satellite is favored over a model without a satellite, the exo-satellite at the given orbital period and mass ratio is detectable. As expected, given a constant RV semi-amplitude, $K$, the mass of the exo-satellite must increase as the semi-major axis increases to remain detectable, following the relationship $K\propto M_{Moon} a^{-1/2}$. 

We find KPIC is sensitive to exo-satellites 2.8\% the mass of GQ Lup B at a separation of \((65 \, R_{\text{Jup}}\)), or the extent of the cavity measured in the CPD. Between the rigid and fluid Roche limits, KPIC is sensitive to 0.6-1.0\% mass ratios. Given our current RV precision, KPIC is much better suited to search for companions with larger mass ratios and to place constraints on satellites likely formed through gravitational instabilities, versus those formed in the CPD.    

\section{Comparison with cavity carving simulations}
\label{sec:carving}

\begin{figure}
  \centering
  \includegraphics[trim={0cm 0cm 0cm 0cm},clip,width=1\linewidth]{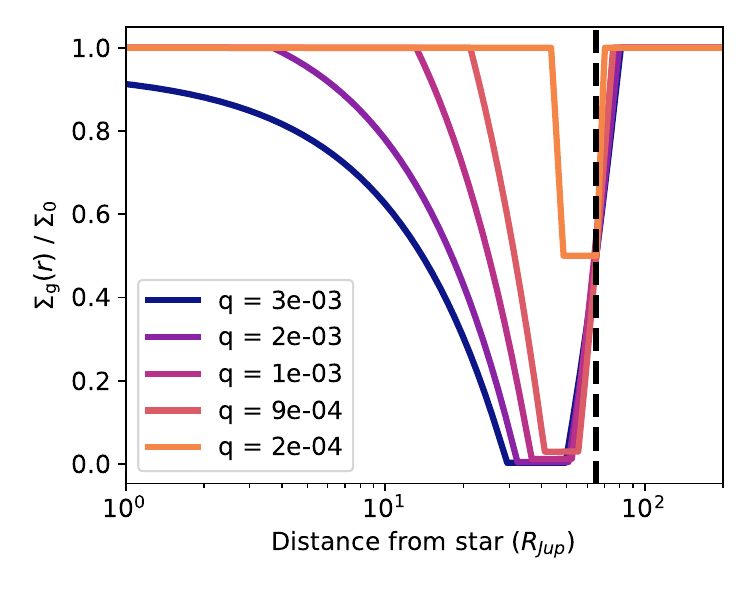}
  \caption{Simulated cavity depth carved by a single satellite. The vertical axis represents the normalized surface density of the CPD, which must be $\leq 0.5$ to produce an prominent cavity. Each colored line represents a different satellite-to-planet mass ratio (q). The blacked dashed line represents the measured radius of the CPD, \(65 \, R_{\text{Jup}}\). We adjust the semi-major axis of the satellite to be consistent with forming a cavity at \(65 \, R_{\text{Jup}}\), regardless of the mass ratio. We find a minimum satellite-to-planet mass ratio of $2 \times 10^{-4}$ is capable of carving a cavity, while the maximum cavity size produced is \(\sim 51 \, R_{\text{Jup}}\).}
  \label{fig:cavity}
\end{figure}

\citet{Stolker2021} fits a disk profile to mid-infrared (MIR) excess found in GQ Lup B observations, determining the temperature and inner radius of the disk to be \(T_{\text{disk}} = 461 \pm 2 \, \text{K}\) and \(R_{\text{disk}} = 65 \pm 1 \, R_{\text{J}}\) respectively. Recent JWST data also suggests that the cavity may be smaller than originally measured, ranging between \(8-40 \, R_{\text{Jup}}\) depending on the model CPD used \citep{Cugno2024}. To investigate the possibility that satellites could create cavities as large as the one observed, we use simulations to estimate the amount of material that different masses of satellites could clear in a CPD. Given that GQ Lup B is actively accreting (\(M \approx 10^{-6.5} M_J \, \text{yr}^{-1}\), \citealp{Stolker2021}) and its CPD is no longer in the decretion phase, we extrapolate physical mechanisms that apply to planets carving out gaps in protoplanetary disks to satellites carving out gaps in circumplanetary disks.

In general, to create cavities in accretion disks surrounding stars, planets excite density waves in the surrounding medium, facilitating the transfer of angular momentum outward. This creates pressure minima and maxima that effectively trap and clear dust. To estimate the amount of material, primarily small dust grains (radii of {\lapprox}0.1 mm), that could be carved out by a single satellite residing in a CPD, we use the prescription to model planetary gap depth outlined in the python package \texttt{DustPy}\footnote{\url{https://stammler.github.io/dustpy/}} \citep{Stammler2022}, following the relations in \citet{Kanagawa2015} modified for the case of a circumplanetary disk instead of a protoplanetary disk,

\begin{equation}
\frac{\Sigma_p}{\Sigma_0}=\frac{1}{1+0.04 K}
\label{eq:surface}
\end{equation}

where

\begin{equation}
K=\left(\frac{M_s}{M_p}\right)^2 h_p^{-5} \alpha^{-1}.
\label{eq:K}
\end{equation}

In \autoref{eq:K}, ${M_s}$ represents the mass of the satellite, ${M_p}$ represents the mass of the planet or brown dwarf, $h_p$ represents the disk aspect ratio, and $\alpha$ represents the viscosity term from the $\alpha$-prescription of kinematic viscosity, $\nu = \alpha c_s h$ \citep{Shakura1973}. To recover the equations above, we consider an axisymmetric, thin disk in two dimensions and embed a satellite within the disk. The satellite exerts a strong gravitational torque on the disk, creating a cavity coinciding with its orbit. \autoref{eq:surface} is an empirical formula relating the gap depth to the mass ratio of the system and disk properties and agrees with observations of cavities carved out by planets in protoplanetary disks \citep{Kanagawa2017}. It does not account for satellite growth or migration.

For our simple estimation of gap depth, the parameter \(K < 1 \times 10^4\) so the disk gap remains in a non-eccentric, steady state. If $K$ becomes too large from increasing the mass ratio between the satellite and companion, \citet{Fung2016} suggests Rayleigh instability may cause filaments of gas and dust to stream into the gap from beyond the edges, thus placing a maximum size on the gap carved. We adopt standard values of \(\alpha = 1 \times 10^{-4}\) and \(h_p = 0.1\) for a common CPD based on \citet{Batygin2020} and vary the satellite-to-planet mass ratio. To maintain consistency with existing literature, we define a cavity as visible when the initial surface density of the disk drops by a factor of 2, thus setting a lower limit on acceptable mass ratios of $q=2 \times 10^{-4}$. Furthermore, we impose the condition \(K < 1 \times 10^4\), establishing an upper limit on accepted mass ratios of $q=3 \times 10^{-3}$ so the cavity remains in a non-eccentric, steady state. Note that the mass ratio corresponding to the largest gap depth is order of magnitude lower than what is currently detectable with KPIC around GQ Lup B. 

\autoref{fig:cavity} illustrates the cavity size and depth that we can expect from mass ratios between $2 \times 10^{-4} < q < 3 \times 10^{-3}$, although this upper limit on the mass ratio does not forbid more massive satellites from forming or creating cavities. Although we see for larger mass ratios that it is possible to carve out a cavity that extends the approximate radius of the one detected in \citet{Stolker2021}, the semi-major axis of the exo-satellite is tuneable parameter in this simulation. We adjust the semi-major axis of the satellite to be consistent with forming a cavity at \(65 \, R_{\text{Jup}}\), regardless of the mass ratio. Given these constraints, we find mass ratios akin to Galilean satellites around Jupiter or larger can carve noticeable cavity sizes, carving a maximum cavity size of \(\sim 51 \, R_{\text{Jup}}\). Although this is shy of the estimated \(65 \, R_{\text{Jup}}\) from \citet{Stolker2021}, it is consistent with recent cavity size estimates from \citet{Cugno2024}. Additionally, our estimation of gap depth does not it does not take into account other physical mechanisms that can also contribute to clearing out the material in the disk closer to the brown dwarf, such as sublimation, magnetic truncation, or the existence of multiple moons. 

\section{Discussion}
\label{sec:Discussion}
\subsection{Prospects with HISPEC}
\label{sec:HISPEC}
Looking to the future, we expect substantial gains in RV precision by using the next generation of high-resolution spectrographs on large telescopes. These gains in RV precision will lead to enhanced sensitivity to systems with lower mass, close in satellites. The High-resolution Infrared Spectrograph for Exoplanet Characterization, Keck/HISPEC, has an expected first light date in late 2026 and will specialize in the high contrast detection and spectroscopy of spatially separated substellar companions \citep{Mawet2019}. With an increased resolution ($R\sim100,000$), wider wavelength coverage (0.98-2.46~$\mu$m), and state of the art calibration techniques facilitated by a laser frequency comb (LFC), HISPEC will be much more sensitive to lower mass companions than current instrumentation. Using the HISPEC Exposure Time Calculator (ETC)\footnote{\url{http://specsim.astro.caltech.edu/hispec_snr}}, we estimate an average RV precision of 22.8 m/s for 180s observations of GQ Lup B. Assuming 18 hours of observation spread across 6 nights, HISPEC achieves the sensitivity shown by the dashed purple line in \autoref{fig:recovery}. HISPEC will increase the expected exo-satellite sensitivity by over an order of magnitude within the measured cavity, making it sensitive to moons that may have formed in the CPD, with mass ratios between $10^{-3} - 10^{-4}$. It will allow access to these seemingly common objects, as outlined in \autoref{sec:occurrence}, marking a significant step forward in our ability to explore and better understand planetary formation. Within the next decade, we expect ELTs to reach RV sensitivity of $\sim$1 m/s for an object like GQ Lup B, translating to a exo-satellite sensitivity (\(\sim10^{-5}\)), exceeding what is needed to detect Galilean satellite analogs \citep{Ruffio2023}. 

\subsection{Occurrence Rates}
\label{sec:occurrence} 
From the theory of how CPDs form satellites, we expect the masses of exo-satellites formed to reflect those of the satellites we see in the solar system, around \(10^{-4}\) of the mass of their host \citep{Canup2006, Sasaki2010}. However, outside of theoretical models and our own solar system, we know little about what to expect. Within the next decade, we can expect Extremely Large Telescopes (ELTs) to be sensitive to Galilean satellites around Jupiter-mass planets, but much sooner, we can expect HISPEC to be sensitive to mass ratios down to \(10^{-4}\), as described in \autoref{sec:HISPEC}. Theories about forming objects with satellite-to-planet mass ratios above \(10^{-4}\) have not been studied in great detail, most likely due to the lack of such larger satellites in the solar system. If planets and satellites formation is governed by similar physics, such as through the aggregation of solid material in disk to form a more massive object \citep{Canup2002, Canup2009, Miguel2016, Ronnet2020, Batygin2020}, it is reasonable to start probing the occurrence rate of this mass regime by extrapolating occurrence ratios for planet-to-star mass ratios and applying them to satellite-to-planet mass ratios. \citet{Suzuki2016} found a peak in the occurrence rate of planets that have a planet-to-star mass ratio between \(10^{-3}\) and \(10^{-4}\) in the MOA-II microlensing survey; however, \citet{Pascucci2018} finds that Kepler data disagrees and argues that this peak in occurrence exists between $10^{-4}$ and $10^{-5}$. HISPEC will be able to probe the former mass ratios, while ELTs will be able to access the latter mass ratios, placing lower mass satellite searches within the capabilities of instruments and facilities in the near future. 

For larger mass ratios, \citet{Lazzoni2024} found that simulations of tidal dissipation during close encounters of massive planets \((1-15 \, M_{\text{Jup}}\)) formed via gravitational instability produced binary planets 14.3\% of the time. If the previous occurrence rates hold true for satellites, they could lend credence to dedicated surveys searching for higher-mass ratio systems, which current and soon-to-be available technologies are sensitive to. 

\section{Conclusion}
\label{sec:conclusion}
In this work, we present the first dedicated RV survey searching for satellites around the directly imaged brown dwarf, GQ Lup B. GQ Lup B stood out as the prime candidate for satellite investigation due to it being the only known companion with a cavity in its CPD \citep{Stolker2021, Cugno2024}, its potential to host moons of comparatively larger masses \citep{Batygin2020}, and its capability for achieving highly precise RV measurements. Although no exo-satellites were found within the cavity of the disk, KPIC is sensitive to exo-satellites 2.8\% the mass of GQ Lup B at a separation of \(65 \, R_{\text{Jup}}\), or the measured radius of the cavity. At the inner most stable point of the cavity, between the rigid and fluid Roche limits, KPIC is sensitive to 0.6-1.0\% mass ratios. 

To explore the feasibility of satellites creating cavities as large as the one observed, we ran simulations to estimate the amount of material that different masses of satellites could clear in a CPD. We expect to see noticeable cavities for mass ratios between $q > 2 \times 10^{-4}$ From our simulations, we find a maximum cavity size of \(\sim 51 \, R_{\text{Jup}}\) carved from a single satellite. Our maximum size estimate does not take into account other mechanisms that can also contribute to clearing out the material in the disk closer to the brown dwarf, such as sublimation, magnetic truncation, or the existence of multiple moons.

Within the next three years, we expect to receive substantial gains in RV precision by using the next generation of high-resolution spectrograph, HISPEC, to continue searching for satellites. The exo-satellite sensitivity of HISPEC will increase by almost two orders of magnitude at close separations, making it sensitive to moons with mass ratios between $10^{-3} - 10^{-4}$. Sensitivity to mass ratios within this range is exciting because it is the expected mass range of satellites forming in the CPD, allowing insights into planet formation and system architecture. Further in the future, the next generation instruments on ELTs are expected to reach RV sensitivity of $\sim$1 m/s for an object like GQ Lup B, translating to a exo-satellite sensitivity (\(\sim10^{-5}\)) great enough to detect solar system moon analogs. Using upcoming high resolution spectrographs to take RV measurements of directly imaged companions is another promising technique that could lead to the discovery of the first exo-satellite.


\begin{acknowledgments}
K.H. is supported by the National Science Foundation Graduate Research Fellowship Program under Grant No. 2139433. 

J.X. is supported by the NASA Future Investigators in NASA Earth and Space Science and Technology (FINESST) award \#80NSSC23K1434.

Funding for KPIC has been provided by the California Institute of Technology, the Jet Propulsion Laboratory, the Heising-Simons Foundation (grants \#2015-129, \#2017-318, \#2019-1312, \#2023-4598), the Simons Foundation, and the NSF under grant AST-1611623.

The W. M. Keck Observatory is operated as a scientific partnership among the California Institute of Technology, the University of California, and NASA. The Keck Observatory was made possible by the generous financial support of the W. M. Keck Foundation. We also wish to recognize the very important cultural role and reverence that the summit of Maunakea has always had within the indigenous Hawaiian community. We are most fortunate to have the opportunity to conduct observations from this mountain. K.H. wishes to acknowledge her settler status on the ancestral lands of the Gabrielino/Tongva people and recognizes that the astronomical observations in this paper were possible because of the dispossession of Maunakea from the Kan\={a}ka Maoli.
\end{acknowledgments}

%

\vspace{5mm}
\facilities{Keck II (KPIC)}


\software{Astropy\footnote{\url{http://www.astropy.org}} \citep{Astropy2013,Astropy2018,Astropy2022}, Matplotlib\footnote{\url{https://matplotlib.org}} \citep{Hunter2007},
PSIsim\footnote{\url{https://github.com/planetarysystemsimager/psisim}},
RVSearch\footnote{\url{https://github.com/California-Planet-Search/rvsearch}} \citep{Rosenthal2021},
KPIC Data Reduction Pipeline\footnote{\url{https://github.com/kpicteam/kpic_pipeline}} \citep{Delorme2021b},
BREADS\footnote{\url{https://github.com/jruffio/breads}} \citep{Ruffio2021, Agrawal2022}
}





\bibliography{bibliography}{}
\bibliographystyle{aasjournal}



\end{CJK*}
\end{document}